\documentclass[english]{emulateapj}
\slugcomment{Accepted for publication in {\it The Astronomical Journal}}

\shorttitle{Ca\thinspace\textsc{ii} Triplet as a Metallicity Indicator}
\shortauthors{Foster et al.}

\makeatletter

\usepackage{graphicx}
\usepackage{amssymb}

\begin{document}

\title{Deriving Metallicities From the Integrated Spectra of Extragalactic Globular Clusters Using the Near-Infrared Calcium Triplet}

\author{Caroline Foster$^{1}$, Duncan A. Forbes$^{1}$, Robert N. Proctor$^{1,2}$, Jay Strader$^{3,4}$, Jean P. Brodie$^{5}$, and Lee R. Spitler$^{1}$}

\affil{$^{1}$Centre for Astrophysics \& Supercomputing, Swinburne University, Hawthorn, VIC 3122, Australia\\
$^{2}$ Universidade de S\~ao Paulo, IAG, Rua do Mato 1226, S\~ao Paulo 05508-900, Brazil\\
$^{3}$Harvard-Smithsonian Center for Astrophysics, 60 Garden St., Cambridge, MA 02138, USA\\
$^{4}$Hubble Fellow\\
$^{5}$UCO/Lick Observatory, University of California, Santa Cruz, CA 95064, USA}

\email{cfoster@astro.swin.edu.au}

\begin{abstract}
The Ca\thinspace\textsc{ii} triplet (CaT) feature in the near-infrared has been employed as a metallicity indicator for individual stars as well as integrated light of Galactic globular clusters (GCs) and galaxies with varying degrees of success, and sometimes puzzling results. Using the DEIMOS multi-object spectrograph on Keck we obtain a sample of 144 integrated light spectra of GCs around the brightest group galaxy NGC 1407 to test whether the CaT index can be used as a metallicity indicator for extragalactic GCs. Different sets of single stellar population models make different predictions for the behavior of the CaT as a function of metallicity. In this work, the metallicities of the GCs around NGC 1407 are obtained from CaT index values using an empirical conversion. The measured CaT/metallicity distributions show unexpected features, the most remarkable being that the brightest red and blue GCs have similar CaT values despite their large difference in mean color. Suggested explanations for this behavior in the NGC 1407 GC system are: 1) the CaT may be affected by a population of hot blue stars, 2) the CaT may saturate earlier than predicted by the models, and/or 3) color may not trace metallicity linearly. Until these possibilities are understood, the use of the CaT as a metallicity indicator for the integrated spectra of extragalactic GCs will remain problematic.

\end{abstract}

\keywords{galaxies: individual: NGC 1407 - galaxies: star clusters - techniques: spectroscopic}

\section{Introduction}
The formation and evolution of galaxies is still an open question. This fundamental problem can be tackled by looking at the ages and abundances of the member stars of nearby galaxies (i.e. galactic archeology). Information about a galaxy's formation and subsequent evolution can be inferred by looking at the spatial distribution and stellar population parameters (e.g.,ages and metallicities) of its constituent stars as they are observed today. This can be done accurately via studying the composition of individual member stars. Unfortunately, except for a handful of very nearby galaxies, it is not possible to resolve individual stars even with the most powerful telescopes available today. One must therefore find means of extracting information about the star formation and enrichment history from the integrated light of stellar populations.

This is easiest done for globular clusters (GCs), which are well approximated as single stellar populations. Using integrated light spectra of GCs together with either empirical relationships or single stellar population (SSP) models, one can estimate stellar population parameters such as ages and metallicities. Moreover, GCs are believed to have formed early in the history of the universe because they are typically measured to be old (e.g.,Forbes \& Forte 2001; Kuntschner et al. 2002; Brodie et al. 2005; Strader et al. 2005; Cenarro et al. 2007, hereafter C07; Proctor et al. 2008). The study of GC systems provides information about the early star formation history of their host galaxies and can be used to constrain galaxy formation scenarios. For example, GC systems typically show a bimodal color distribution. Although there has been some debate on its origin (Yoon, Yi \& Lee 2006), the currently favored interpretation is that the bimodality in color reflects a metallicity bimodality because of the predominately old ages of GCs (Brodie \& Strader 2006). Therefore, the blue and red subpopulations correspond to a metal-poor and metal-rich subpopulation, respectively. Any galaxy formation scenario must provide at least two different star formation epochs or mechanisms to account for the metallicity bimodality of the host's GC system (but see Muratov \& Gnedin 2010).

Early-type galaxies typically have large numbers of GCs and are thus an excellent target for GC studies. For this reason, the present study concentrates on the elliptical galaxy NGC 1407, a brightest group galaxy (BGG) dominating the Eridanus A group (Brough et al. 2006). It harbours a rich GC system (Perrett et al. 1997; Harris et al. 2006; Forbes et al. 2006; Spitler et al. 2010, in preparation). As with other BGGs and brightest cluster galaxies (e.g.,Harris 2009a), NGC 1407's blue GC subpopulation shows a trend of color with luminosity such that brighter GCs have redder colors on average. This `blue tilt' has been observed in several, usually massive, galaxies of different morphological types and is usually interpreted as a mass-metallicity relationship (see Strader et al. 2006; Harris et al. 2006; Mieske et al. 2006; Spitler et al. 2006; Strader \& Smith 2008; Bailin \& Harris 2009; Forbes et al. 2010; Blakeslee et al. 2010). Although this interpretation has been questioned by some (Kundu 2008; Waters et al. 2009), its reality has been confirmed by Harris (2009b) and Peng et al. (2009).

In order to assess the formation history, enrichment history, confirm the origin of the color bimodality and the blue tilt in GC systems, statistically significant samples of spectroscopically determined metallicities of GCs are required. Spectroscopic data of large GC samples are rare and typically time consuming to acquire. However, using the DEIMOS multi-object spectrograph on Keck it is possible to obtain over 100 integrated light spectra of GCs simultaneously. Using DEIMOS, we have obtained over 100 spectra of kinematically confirmed GCs around NGC 1407 suitable for stellar population analysis.

The sensitivity of DEIMOS is good at red wavelengths near the region of the Ca\thinspace\textsc{ii} triplet (hereafter CaT, 8498, 8542, and 8662 \AA). The CaT is known to correlate with metallicity for integrated light spectroscopy of Galactic GCs (Bica \& Alloin 1987; Armandroff \& Zinn 1988). The CaT has also been studied as a potential metallicity indicator for integrated light of \emph{galaxies} in the past with varying degrees of success. The word `puzzle' has been put forward with regards to its `unexpected' behavior with metallicity. For example, the CaT was found to be lower than predicted by theory in giant elliptical galaxies (Saglia et al. 2002) and higher than predicted in dwarf elliptical galaxies (Michielsen et al. 2003). Possible resolution of the puzzle has been obtained by comparing CaT strengths to metallicities determined using optical spectra in dwarf ellipticals rather than metallicities derived from narrow-band photometry (Michielsen et al. 2007). Nevertheless, Cenarro, Cardiel \& Gorgas (2008) and Foster et al. (2009) were able to successfully use the CaT to probe the metallicity gradients of M32 and a sample of massive to intermediate mass elliptical galaxies, respectively. Therefore, while the behavior of the CaT with respect to metallicity has been studied for the integrated light spectra of galaxies, which are composite stellar populations, with varying degrees of success, it is worth investigating whether it can be used straightforwardly as a metallicity indicator for the integrated light of simple stellar populations such as extragalactic GCs.

A description of the data used in this work is presented in Section \ref{sec:data}. In Section \ref{sec:analysis} we analyze the observational and theoretical behavior of the CaT with metallicity and present our choice of CaT index definition. Results can be found in Section \ref{sec:results}. Finally, a discussion and our conclusions are given in Sections \ref{sec:discussion} and \ref{sec:conclusions}, respectively.

\section{Data}\label{sec:data}
	\subsection{Photometry}

The photometric data consist of imaging of the central region (3.4 x 3.4 arcmin) from the Advanced Camera for Surveys (ACS) mounted on the \emph{Hubble Space Telescope} (HST) with both the $F435W$ ($B$) and $F814W$ ($I$) bands. The ACS dataset has been independently analyzed and published by both Forbes et al. (2006) and Harris et al. (2006). Here this is supplemented by Subaru/Suprime-Cam images covering a wider field of view (34 x 27 arcmin) in the SDSS $g$, $r$, and $i$ filters (see Spitler et al. 2010, in preparation). Globular cluster candidates in the central region imaged by both Suprime-Cam and ACS thus have photometry in the $B$, $I$, $g$, $r$, and $i$ filters while those outside the ACS field only have $g$-, $r$-, and $i$-band photometry. In all cases, the reddening corrections were performed according to the DIRBE dust maps (Schlegel, Finkbeiner \& Davis 1998).

\begin{figure}
\epsscale{1.19}
\plotone{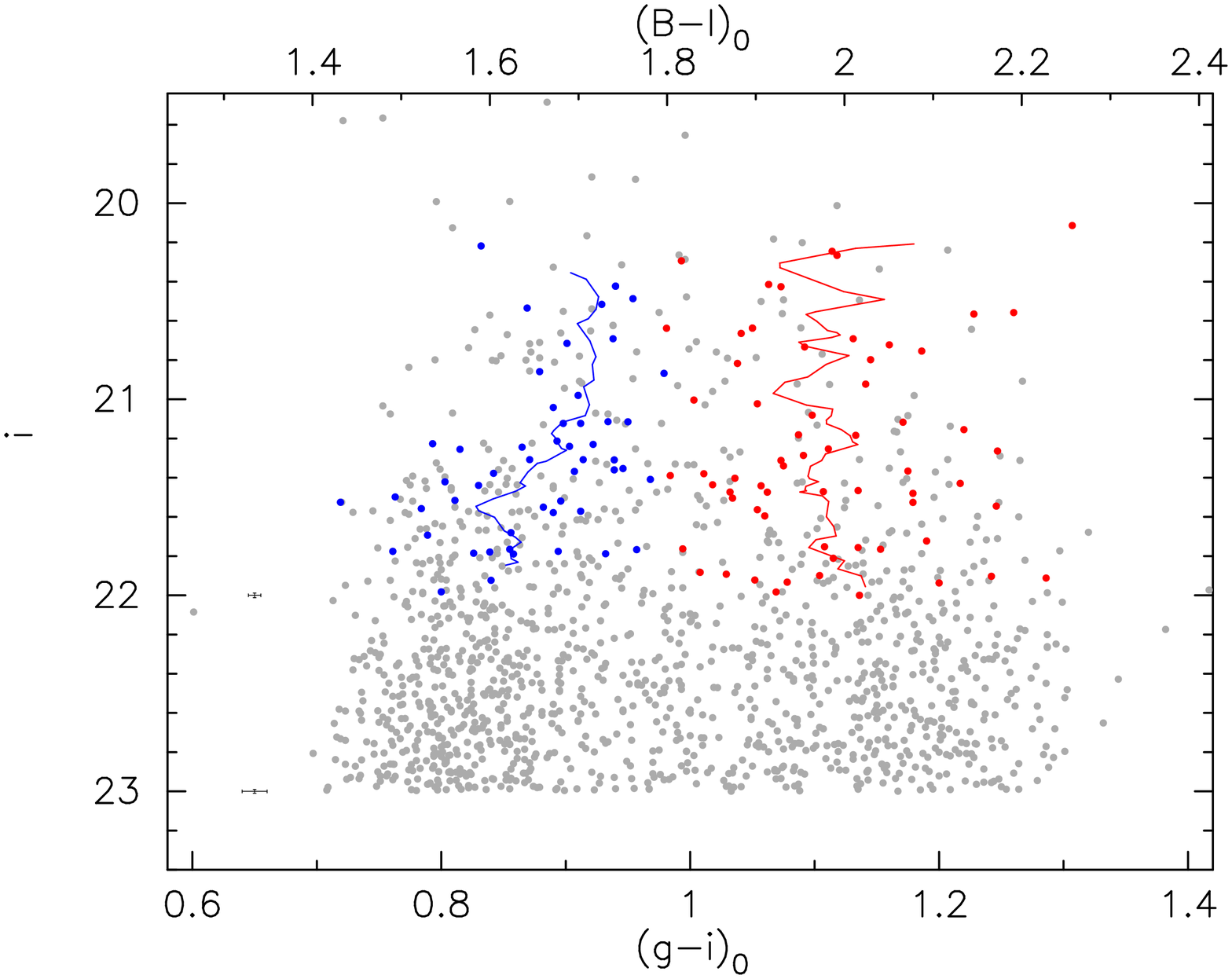}
\caption{Colour magnitude diagram for NGC 1407 GCs brighter than $i=23.0$ from Subaru/Suprime-Cam data. Grey points show the position of all photometrically selected GC candidates (including a small number of potential NGC 1400 GCs). Red and blue points show the spectroscopically confirmed GCs with CaT measurements. A running average (solid blue and red lines) of our sample has been overplotted for both GC subpopulations indicating a blue tilt but no red tilt. The color division is $(g-i)_{0} = 0.93$ following Romanowsky et al. (2008). The top x-axis shows the color conversion which we derive in Figure \ref{fig:color_conv} for comparison. }\label{fig:cmd}
\end{figure}

Both Forbes et al. (2006) and Harris et al. (2006) find a relationship between color and luminosity for the blue GC subpopulation (blue tilt) in their ACS imaging of NGC 1407. Kundu (2008) and Waters et al. (2009) have argued that the blue tilt in M87 (also a massive elliptical galaxy) could be a photometric bias caused by the resolved sizes of the brightest metal-poor GCs in the HST/ACS images as opposed to an intrinsic astrophysical phenomenon. However, these claims have recently been refuted by both Harris (2009b) and Peng et al. (2009). The Subaru/Suprime-Cam color magnitude diagram (CMD) for all photometrically selected GC candidates brighter than $i=23.0$ around NGC 1407 is shown in Figure \ref{fig:cmd}. This apparent magnitude limit corresponds to an absolute magnitude of $M_i=-8.9$ when assuming the average redshift independent distance modulus $(m-M)=31.9$ given in NED\footnote{NASA/IPAC Extragalactic Database (NED) is operated by the Jet Propulsion Laboratory, California Institute of Technology, under contract with the National Aeronautics and Space Administration.}. The CMD reveals a bimodal GC distribution as found by Forbes et al. (2006) and Harris et al. (2006). The photometric blue tilt is apparent and there is no equivalent red tilt. We also show our spectroscopic subsample and compute the running average for both the red and blue subpopulations. The distribution of our selected spectroscopic subsample in the CMD is representative of the underlying GC color distribution for $i\le22.0$ (or $M_i\le-9.9$).

\begin{figure}
\epsscale{1.19}
\plotone{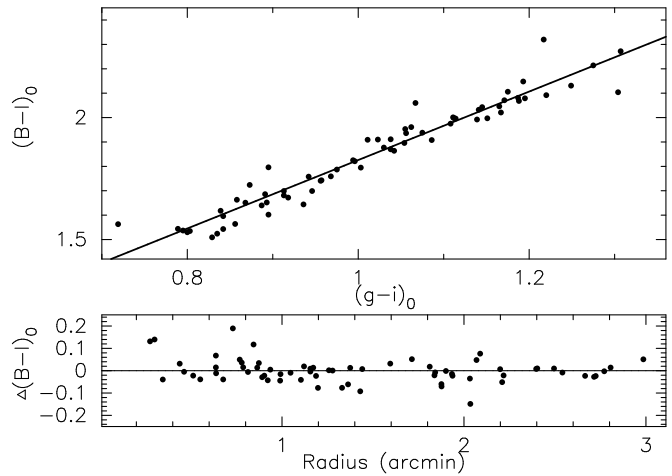}
\caption{Empirical color conversion from $(g-i)_{0}$ to $(B-I)_{0}$. The top panel shows the least squares linear fit to the data. One outlier due to crowding was removed from the fit and is not shown. Differences between converted and actual $(B-I)_{0}$ colors with galactocentric radius are shown underneath. These residuals decrease with radius down to a few hundredth of a mag for radii $\gtrsim1.0$ arcmin.}\label{fig:color_conv}
\end{figure}

In order to obtain $(B-I)_{0}$ colors and \emph{combine} both datasets (i.e. HST/ACS and Subaru/Suprime-Cam) we have used the candidates in the common central region to derive the following empirical conversion between $(g-i)_{0}$ and $(B-I)_{0}$ colors:
\begin{equation}
(B-I)_{0}=(1.40\pm0.05)(g-i)_{0}+(0.49\pm0.04).
\end{equation}
This conversion is useful when comparison our results to single stellar population models that do not always provide SDSS colors (see Section \ref{sec:models}). Figure \ref{fig:color_conv} shows the color conversion as well as the residuals with projected galactocentric radius. The residuals are slightly higher at small projected galactocentric radii due to crowding and the more uncertain Subaru photometry near the center where NGC 1407's surface brightness is high. However, as we move to larger projected radii the error in the converted $(B-I)_{0}$ is reduced to only a few hundredth of a mag (see the lower panel of Figure \ref{fig:color_conv}).

\begin{figure}
\epsscale{1.19}
\plotone{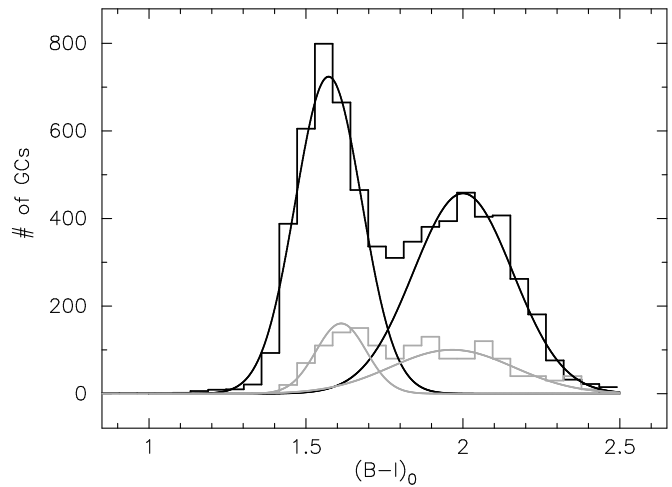}
\caption{Colour histogram of our combined (Subaru/Suprime-Cam and HST/ACS) photometric data for all GC candidates with $i\le23.0$ (black) and our spectroscopic subsample scaled by a factor of 10 for clarity (grey). Overlaid are the Gaussian profiles estimated by the KMM test (assuming heteroscedasticity) with mean blue and red peaks at $(B-I)_{0}=1.57$ and 2.00, respectively, for the whole photometric sample.}\label{fig:BIhist}
\end{figure}

In Figure \ref{fig:BIhist} we show both the color histogram of our combined data and the spectroscopic subsample, whose color distribution is representative of that of the combined data. In order to directly compare with previous literature, we apply the heteroscedastic (unequal widths) KMM test to all candidates within galactocentric radii $\le2$ arcmin (comparable to the ACS field). This yields $(B-I)_{0}=1.62$ and 2.07 for the mean color of the blue and red GC subpopulations, respectively. These values are in good agreement with the previous comparable analyses based on ACS imaging of Harris et al. (2006) and Forbes et al. (2006) who found blue peaks at 1.63 and 1.61, and red peaks at 2.07 and 2.06, respectively. However, because of radial color gradients similar to those found by Harris (2009b) in M87, the larger Subaru field of view has peaks at $(B-I)_0=1.57$ and 2.00 with widths of 0.15 and 0.16 for the blue and red GC subpopulations around NGC 1407, respectively. Using the SSP models of Vazdekis et al. (2003, hereafter V03) the difference in the average $(B-I)_0$ colors for the two subpopulations translates into a metallicity difference of $\Delta$[Fe/H]$\approx1.0$ for a fixed old age ($\sim13$ Gyr) or an age difference of $> 8$ Gyr assuming a fixed moderate metallicity of [Fe/H]$=-0.38$ for all GCs. The latter can be ruled out because an age difference as large as 8 Gyr would have already been detected spectroscopically (see C07). We refer the reader to Spitler et al. (2010, in preparation) for further details on the photometry.

	\subsection{Spectroscopy}
		\subsubsection{Acquisition}
The Keck/DEIMOS spectroscopic data of our photometrically selected GC candidates was obtained during two distinct observing runs.

First, three masks were observed on the nights of 2006 November 19-21. The 1200 l mm$^{-1}$ grating with 7500 \AA\space central wavelength and 1 arcsec slit width was used. Details of the first observing run can be found in Romanowsky et al. (2009). The second observing run occurred on the nights of the 2007 November 12-14 under good seeing conditions (typically $\sim 0.7$ arcsec). Three masks were observed with similar instrumental setup as that of Romanowsky et al. (2009). The 1200 l mm$^{-1}$ grating centred on 7800 \AA\space was used. The setup for both observing runs allows coverage of the wavelength range $\sim7550-8900$ \AA\space at redder wavelengths with a resolution of $\sim1.5$\AA\space around the CaT features. Using this setup, we also occasionally cover H$_{\rm \alpha}$ (6563\AA) at bluer wavelengths for a small fraction of our objects. Four 30 minute exposures were taken on each mask, yielding a total exposure time of 2 hours.

		\subsubsection{Data reduction}\label{sec:reduction}
\begin{figure}
\epsscale{1.19}
\plotone{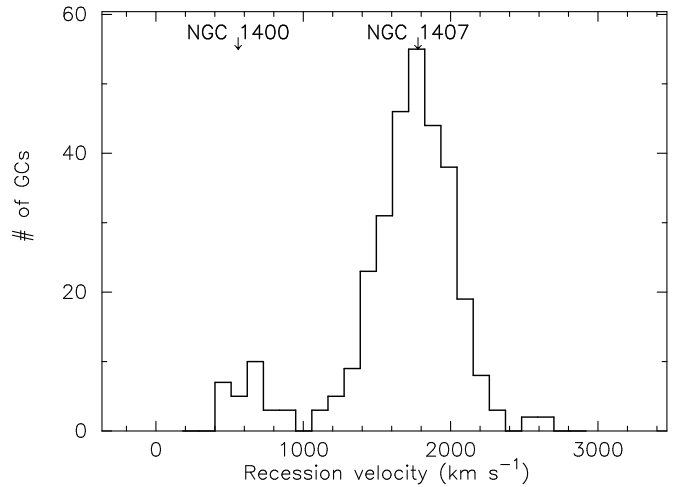}
\caption{Histogram showing the recession velocity of all confirmed GCs around NGC 1400 and NGC 1407. The systemic velocities of NGC 1400 and NGC 1407 are labelled to show that their respective GC systems are easily delineated in velocity space with a gap at $\approx1000$ km s$^{-1}$.}\label{fig:rvhist}
\end{figure}

The reduction of the 2006 data is described in Romanowsky et al. (2009). As with the 2006 data, the \textsc{idl} spec2d data reduction pipeline written for the DEEP2 galaxy survey was used to reduce the 2007 DEIMOS data. The spectra were then cross-correlated with the solar spectrum to extract radial velocities using the \textsc{fxcor} routine in \textsc{iraf}\footnote{\textsc{iraf} is distributed by the National Optical Astronomy Observatories, which are operated by the Association of Universities for Research in Astronomy, Inc., under the cooperative agreement with the National Science Foundation.}. Using these velocities, it was possible to clearly distinguish between GCs belonging to NGC 1407, background unresolved galaxies and foreground stars. One mask included some NGC 1400 GCs, but those were easily identified because of the significantly different recession velocity of NGC 1400 (558 km s$^{-1}$, NED) and its GC system from that of NGC 1407 (1779 km s$^{-1}$, NED) as seen from Figure \ref{fig:rvhist} (see also Romanowsky et al. 2009). The velocity selection yielded 113 new confirmed GCs around NGC 1407. These data were supplemented by those of Romanowsky et al. (2009) to give a total of 274 distinct confirmed GCs. The radial velocities of the 7 objects in common between both the 2006 and 2007 data show very good agreement with an rms scatter of only 11 km s$^{-1}$.

Because the NIR region of the spectrum is strongly affected by skylines we find that residual skylines in our raw spectra leftover from the sky subtraction influence our index measurements. In order to counteract this, the spectra are template fitted with the \textsc{pPXF} code of Capellari \& Emsellem (2004) using 13 stellar templates that were observed during the 2007 November run with the same instrumental setup as the GC data. The templates include 11 giant and 2 dwarf stars spanning spectral types from F to early M and a wide range in CaT depth. Known problematic regions of the spectrum with strong skylines were excluded during the fitting procedure (see Proctor et al. 2009; Foster et al. 2009). The \textsc{pPXF} routine redshifts, broadens, and chooses the weighted combination of the templates that minimizes the residuals between the raw spectrum and the fit. The routine is unable to fit the noisiest spectra as well as the incomplete spectra affected by vignetting due to their position near the edges of the DEIMOS mask. These $\sim 20$ spectra are not used further for index measurements. The final \textsc{pPXF} fitted spectra are the ``best'' skyline residual free description of our spectra assuming only minimal template mismatches. This technique is similar, and more accurate in principle, than fitting purely Gaussian profiles to the CaT features (Battaglia et al. 2008). 

\begin{figure}
\epsscale{1.19}
\plotone{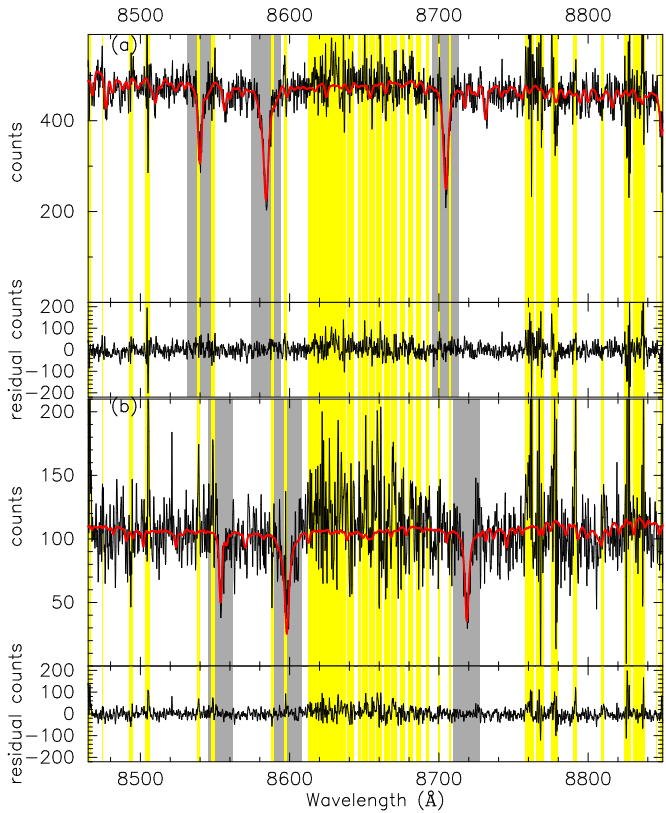}
\caption{Panels (a) and (b) show the raw spectra of a sample bright ($i=20.20$) and a faint ($i=21.98$) GC, respectively. Overlaid in red are the fitted spectra. Residuals, mostly caused by skylines (yellow highlight) and noise, are shown underneath for each fit. Regions shaded in grey show the CaT index positions.}\label{fig:templates}
\end{figure}

Figure \ref{fig:templates} shows examples of template fitted spectra for one of the brightest and one of the faintest GCs in our final sample. The residuals are mostly uniform but showing some features that are mostly associated with the position of known skylines, indicating that template mismatch is not significant. Next, the best fit spectra are continuum normalized interactively using the \textsc{iraf} \textsc{continuum} routine with a spline3 function of order typically $\sim 4$ and a higher and (stricter) lower sigma clipping to ensure that spectral features are not fitted. This sets the continuum to unity. In what follows, we will refer to these \textsc{pPXF} and continuum normalized spectra as ``fitted spectra'' in order to distinguish them from the ``raw spectra'' output from the DEIMOS pipeline. Finally, the fitted spectra with a raw average number of counts less than 80 were removed from the sample. This corresponds to an apparent/absolute $i$-band magnitude and signal-to-noise ratio (S/N) cut of approximately 22.0/--9.9 mag and 9 per \AA, respectively (see Appendix \ref{Appendix:errors}). The final sample contains 144 GCs associated with NGC 1407.

The index values are measured on both the fitted and raw spectra. Determination of the error on the index values is discussed in Appendix \ref{Appendix:errors}.

\section{Analysis}\label{sec:analysis}
In this section, we first review the observational evidence for the sensitivity of the CaT to metallicity in the integrated light spectra of GCs. We then describe and motivate our choice of CaT index definition. We apply our index definition to model spectra from V03 and Bruzual \& Charlot (2003, hereafter BC03) in order to better understand theoretically the behavior of the CaT with metallicity. These will form the observational and theoretical bases on which our results are obtained and discussed.

\subsection{Revisiting Armandroff \& Zinn (1988)}\label{sec:AZ88}

\begin{figure}
\epsscale{1.19}
\plotone{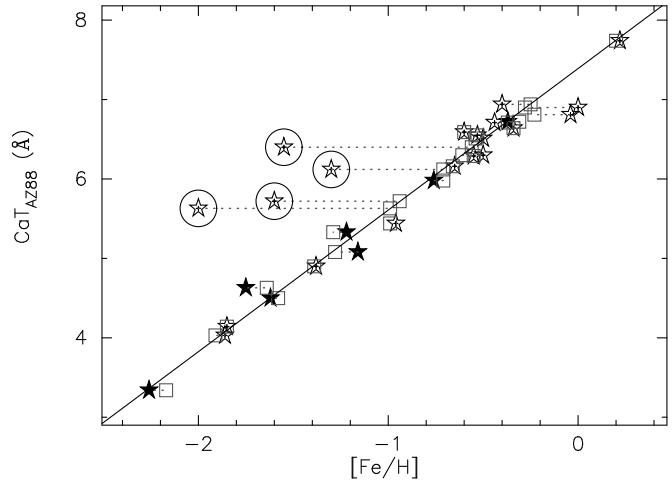}
\caption{An updated reproduction of Figure 5 of AZ88 showing the tight correlation between CaT$_\mathrm{AZ88}$ and [Fe/H]. CaT$_\mathrm{AZ88}$ index values are taken directly from AZ88. Hollow squares are the metallicity measurements quoted by AZ88. Stars represent updated [Fe/H] values from H96 (2003 update) with filled symbols showing the 7 points chosen by AZ88 to determine their relationship (solid line). Dotted lines are drawn between the corresponding measurements for a given GC to guide the eye. The four circled outliers are discussed in the text. Recent metallicity measurements have left the original relationship found by AZ88 essentially unchanged for [Fe/H]$\lesssim-0.4$.}\label{fig:AZ88}
\end{figure}

The CaT was first recognised as a potential metallicity indicator in the works of Bica \& Alloin (1987) and Armandroff \& Zinn (1988, hereafter AZ88). In particular, the latter obtained integrated spectra of Galactic GCs to measure the CaT index and averaged literature metallicity measurements. After removing the data points with uncertain metallicity measurements they used linear regression to fit a straight line to the remaining 7 data points (i.e. 47 Tuc, NGC 362, NGC 1851, NGC 5927, NGC 6093, M15, and M2). They obtained the linear relationship:
\begin{equation}
\mathrm{[Fe/H]}=-4.146 + 0.561 \times \mathrm{CaT_{AZ88}}, \label{eq:AZ88}
\end{equation}
with rms scatter of only 0.12 dex. This relationship is shown in Figure \ref{fig:AZ88}. The metallicities are the average between the quoted values in the literature at the time and their CaT$_\mathrm{AZ88}$ derived values. There exists more recent metallicity measurements for several GCs and it is worthwhile to verify that the then observed tight correlation still holds with these newer data. In Figure \ref{fig:AZ88} the new measurements as compiled in the 2003 updated version of the Catalogue of Parameters for Milky Way Globular Clusters (Harris 1996, hereafter H96) are also shown. In what follows, we use [Fe/H] to denote metallicities although the metallicities given in both AZ88 and H96 (and hence herein) are based on the Zinn \& West (1884) metallicity scale, which is not a strict Fe scale (e.g.,Carreta \& Gratton 1997; Rutledge, Hesser \& Stetson 1997).

Immediately striking in Figure \ref{fig:AZ88} are the four circled data points. These are four of the eight GCs for which AZ88 were the first to give a metallicity estimate. The GCs are HP 1 (Ortolani, Bica \& Barbuy  1997a), Terzan 1 (Ortolani et al. 1999a), Terzan 4 (Ortolani, Barbuy \& Bica 1997b; Origlia \& Rich 2004), and Terzan 9 (Ortolani et al. 1999b). All four are bulge GCs, therefore estimates of their metallicity are plagued by extinction and contamination by foreground bulge stars. In fact, as explained in the above respective references, it is possible that the CaT$_\mathrm{AZ88}$ measurements were contaminated by foreground metal-rich bulge stars. More particularly, as explained in Barbuy et al. (2006), the metallicity of HP1 is a matter of debate with metallicity estimates differing by as much as 1.2 dex. In any case, these four uncertain data points were not originally used by AZ88, leaving the relatively tight relationship unaltered.

A linear fit to the updated data for the original 7 GCs selected by AZ88 (filled stars in Figure \ref{fig:AZ88}) yields results consistent with that of AZ88 within $1\sigma$, indicating that more recent metallicity measurements have left the relationship unchanged. However, the most metal-rich GC (NGC 5927) used for this analysis has a metallicity of $-0.4$ dex and it is unclear whether the relationship remains linear beyond this point. Nevertheless, this confirms that it is still justified to obtain metallicities for Galactic GCs from the CaT$_\mathrm{AZ88}$ index measurements using the AZ88 relationship at least for [Fe/H]$\lesssim-0.4$ and possibly beyond.

\subsection{Choice of index definition}\label{sec:IndexDef}

Several definitions for the CaT index have been used in the literature. Among the definitions that apply to integrated spectroscopy are those of AZ88; Diaz, Terlevich \& Terlevich (1989, hereafter DTT89); and Cenarro et al. (2001, hereafter C01). For a review covering these index definitions and how they compare see C01. As mentioned above, AZ88 used integrated spectra of GCs to measure the CaT index. Their definition (CaT$_\mathrm{AZ88}$) therefore should be suitable for GC integrated light spectra such as the present dataset. The CaT index definition of DTT89 (CaT$_\mathrm{DTT89}$) is slightly broader, thus more suitable for the study of galaxy integrated light for which velocity dispersion broadens the CaT features. One drawback of this definition is that it uses the same two continuum passbands for all three CaT lines and is thus strongly affected by changes in the shape of the continuum. Because hot stars (B, A, and F types) have pronounced Paschen lines, three of which overlap with the CaT, one has to worry about contamination of the CaT by Paschen lines from such stars. Therefore, when studying the behavior of the CaT in a wide range of stellar temperatures the C01 definition (CaT$^{*}_\mathrm{C01}$) may be preferable because it is designed to correct for the Paschen line contribution.

Provided that the GCs in NGC 1407 are comparable to their Galactic counterparts, one can use the empirical correlation obtained by AZ88 to obtain metallicities. The CaT$_\mathrm{AZ88}$ index central passbands ($Ca1 =$ [8490.0-8506.0\space\AA$]$, $Ca2 =$ [8532.0-8552.0\space\AA], $Ca3 =$ [8653.0-8671.0\space\AA]) were thus adopted for this work. As the typical velocity dispersion of GCs is small, the relatively narrow definition of AZ88 is appropriate. Also, contamination by Paschen lines should not be a concern for integrated spectroscopy of GCs because they are strong only in the theoretical spectra of young stellar populations ($\lesssim 2$ Gyrs, but see Appendix \ref{sec:SmallLines}). Therefore, the passbands of the index definition of AZ88 were selected in order to use their empirical conversion to metallicity.

\begin{figure}
\epsscale{1.19}
\plotone{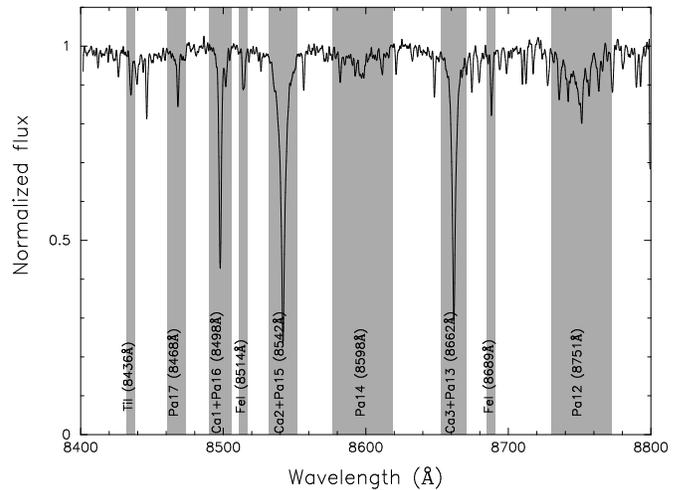}
\caption{Example of a fitted spectrum with the main spectral features highlighted. Identified spectral features include the CaT definition (Pa13, Pa15, and Pa16 lines overlap with the Ca3, Ca2, and Ca1, respectively), and the C01 Paschen `triplet' lines (Pa17, Pa14, and Pa12). Three other metallicity sensitive lines are also highlighted (see Appendix \ref{sec:SmallLines}).}\label{fig:indices}
\end{figure}

\begin{figure}
\epsscale{1.19}
\plotone{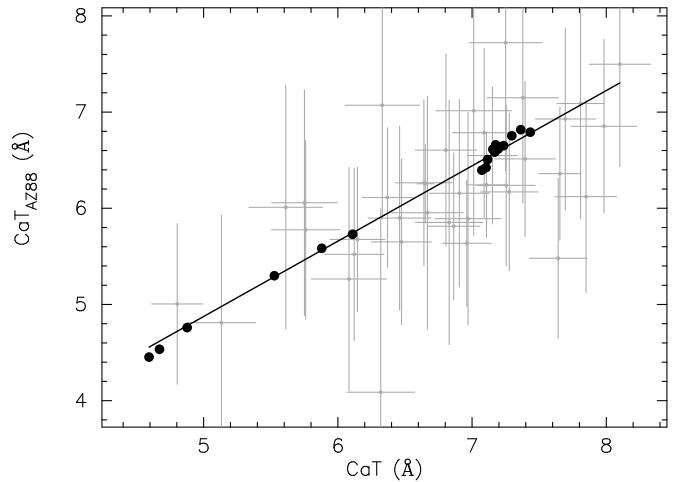}
\caption{An assessment of the systematics introduced by our index measurement method. The x-axis shows the CaT index value measured on the fitted spectra while the y-axis shows the same value measured using the AZ88 index definition on the raw spectra. Black and gray points show values measured using old ($>8$ Gyr) SSP model spectra from V03 and our GC spectra with S/N$\ge15$, respectively. The solid line shows the best fit line (Equation \ref{eq:convCaT}) through the model points that is used to correct our index measurements.}\label{fig:compare}
\end{figure}

In Figure \ref{fig:indices} we identify some of the features present in the CaT region of the spectrum. Our adopted definition of the CaT index is shown. Because the fitted spectra are continuum normalized, the continuum was set to unity to compute the fitted indices and the continuum passbands of AZ88 were not needed. The CaT indices computed using this method will be referred to simply as CaT in what follows. The continuum normalization has introduced some systematic differences between our index values and those of AZ88. Indeed, like flux calibration, continuum normalization can cause systematic deviations that are difficult to quantify precisely without access to the original AZ88 data (see C01 for a discussion of this). In order to quantify the systematics we measure both the CaT$_\mathrm{AZ88}$ (i.e. measured on the raw spectra) and CaT (i.e. after fitting) indices on the SSP models of Vazdekis et al. (2003) for ages of 8 Gyrs and older. Figure \ref{fig:compare} shows the comparison, which has an offset but a very small scatter. The same is true for our GC spectra, albeit with larger scatter due mostly to the effects of skyline residuals on the measured CaT$_\mathrm{AZ88}$. The equation of the best-fit line to the V03 SSP models points is:
\begin{equation}
\mathrm{CaT_{AZ88}}=(0.783\pm0.025)\times \mathrm{CaT}+(0.96\pm0.17), \label{eq:convCaT}
\end{equation}
with $r^2=0.97$ implying that 97 \% of the V03 data are accounted for by the derived relationship. Combining with Equation \ref{eq:AZ88} above yields the following conversion between CaT and [Fe/H]:
\begin{equation}
\mathrm{[Fe/H]_{CaT}}=-3.641+0.438\times \mathrm{CaT}.\label{eq:convtoFe}
\end{equation}
We will use Equation \ref{eq:convtoFe} to convert our CaT measurements into metallicities. Another potential source of systematics between our measured indices and those of AZ88 is the difference in resolution (corresponding to $\sigma=51$ km s$^{-1}$ in our case). Fortunately, as shown in Figure 5b of C01, the relative measurement error induced by such a difference in resolution of $\sigma=51$ km s$^{-1}$ for the CaT$_\mathrm{AZ88}$ index is insignificant.

\subsection{Single stellar population models}\label{sec:models}

SSP models can provide theoretical insight for understanding the behavior of the CaT index. Indeed, GCs are believed to be well approximated as SSPs. One should thus be able to directly compare GC properties with those of SSP models in the literature. Unfortunately, as will be demonstrated below, different SSP model sets make discrepant predictions about the behavior of the CaT index with metallicity.

An advantage of using SSP models, particularly those providing spectral energy distributions (SEDs), is that it is possible to perform template fits and continuum normalization in order to measure the CaT index so as to self-consistently directly compare the observed measurements with the model predictions. The present analysis is compared to the V03 and the BC03 SSP models, which both supply SEDs of sufficient resolution.

Moreover, the V03 and BC03 models include photometry in the Johnson-Cousins system and a wide range of metallicities and ages. We use the V03 SSP models with a Kroupa (2001) initial mass function (IMF) with metallicities between $-1.68\le$[Fe/H]$\le0.02$. For the BC03 models, we use the SSPs with a Chabrier (see Chabrier 2003) IMF (similar to the Kroupa IMF) in the metallicity range $-1.7\le$[Fe/H]$\le0.00$ so as to directly compare with the V03 models. Models with ages of 5, 9, and 13 Gyrs are selected in both V03 and BC03.

\begin{figure}
\epsscale{1.19}
\plotone{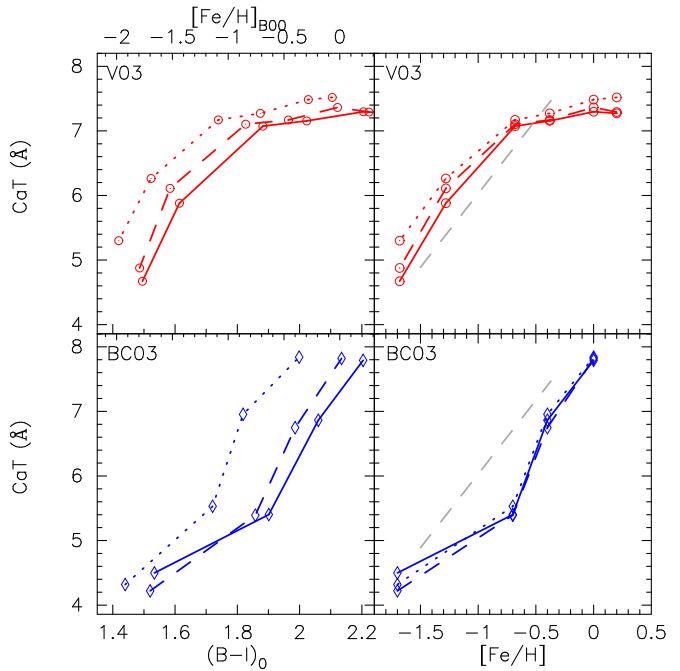}
\caption{Predictions from V03 and BC03 SSP models. The dotted, dashed, and solid lines correspond to 5, 9, and 13 Gyrs models. From left to right, the red hollow circles on the V03 model lines correspond to [Fe/H]$= -1.68, -1.28, -0.68, -0.38, 0.00, 0.02$ and the blue hollow diamonds on the BC03 model lines correspond to [Fe/H]$= -1.7, -0.7, -0.4, 0.0$. The top x-axis of the left panels shows the [Fe/H]-values for the conversion from $(B-I)_{0}$ by B00 based on Galactic GCs. Our conversion between CaT and $[FeH]$ for [Fe/H]$\le-0.4$ (Equation \ref{eq:convtoFe}) is shown as a dashed gray line for comparison.}\label{fig:models}
\end{figure}

Figure \ref{fig:models} shows the predictions of the V03 and BC03 models for CaT versus $(B-I)_{0}$ and metallicity. When no other independent metallicity measurement is available, colors are used as a diagnostic for the sensitivity of the CaT to metallicity. The $(B-I)_{0}$ colors of GCs are known to correlate with metallicity for Galactic and extragalactic GCs (see for example Barmby et al. 2000, hereafter B00; Spitler, Forbes \& Beasley 2008). However, as shown by Smith \& Strader (2007), this conversion is affected by the selected Galactic GC sample from which it is derived due to observational errors and/or interstellar reddening.

From Figure \ref{fig:models}, we see that there is broad agreement between the V03 and BC03 SSPs in the overall range of the CaT values and its qualitative sensitivity to metallicity, though they differ greatly in the details. Indeed, the two different sets of models assign widely different metallicities to the same CaT value. A feature that is found in the V03 models is the `loss of sensitivity' or `saturation' of the CaT features to metallicity as the metallicity increases beyond [Fe/H]$\sim$[Fe/H]$_{\mathrm{CaT}}\sim-0.5$ dex for the $\sim13$ Gyrs model. This qualitative behavior is in disagreement with that predicted by the BC03 models for which the CaT sensitivity to metallicity appears to increase for higher values of [Fe/H].

The effect of age on the measured CaT index is minimal for old ages (e.g.,DTT89; V03; Carrera et al. 2007). Indeed, as can be seen on Figure \ref{fig:models} both the V03 and BC03 models show very small variations of the CaT for ages $\gtrsim5$ Gyrs.

The empirical calibration of AZ88 (Equation \ref{eq:AZ88}), which is based on Galactic GCs, appears to lie below and above the V03 and BC03 models, respectively, in Figure \ref{fig:models}. We note however that the slope of the V03 models at low metallicities ([Fe/H]$\lesssim-0.5$) is in good agreement with Equation \ref{eq:convtoFe}. This was already shown in Figure 14 of V03.

In summary, we find that the BC03 and V03 models make discrepant predictions with respect to the behavior of the CaT as a function of metallicity. For this reason, we choose to use the empirically derived conversion of AZ88 in this work.

\section{Results: The GC system of NGC 1407}\label{sec:results}

\begin{figure}
\epsscale{1.19}
\plotone{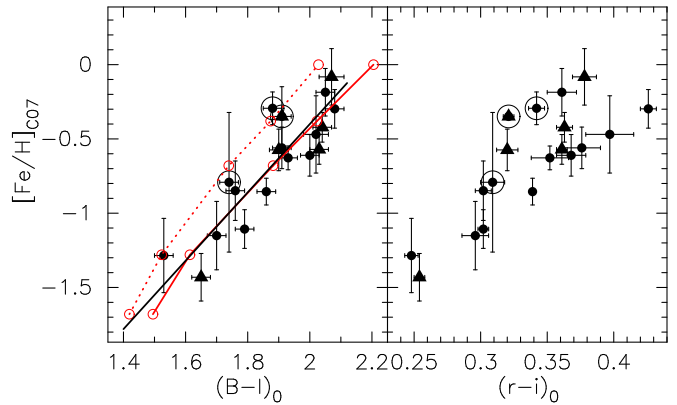}
\caption{Relationship between [Fe/H] and color for a sample of NGC 1407 GCs based on blue spectroscopic data from C07. The hollow circles highlight the 3 GCs identified in C07 as potentially young or harbouring BHBs while the triangles show common GCs between this work and that of C07. The solid black line shows the B00 relationship while the solid and dotted red lines represent the 13 and 5 Gyrs V03 model predictions, respectively.}\label{fig:C07data}
\end{figure}

NGC 1407 is a brightest group galaxy (BGG) showing clear GC color bimodality with a division between the blue (metal-poor) and red (metal-rich) GCs occurring around $(B-I)_{0}=1.84$ or $(g-i)_{0}=0.93$ (see Harris et al. 2006; Forbes et al. 2006; Romanowsky et al. 2009). C07 obtained Keck/LRIS spectra of the brightest GC candidates. They derived metallicities and ages for 19 confirmed GCs and 1 ultra compact dwarf (UCD) using the Lick/IDS system (Gorgas et al. 1993; Worthey et al. 1994) and the method of Proctor, Forbes \& Beasley (2004). They found the majority to be old, with 3 being either young (i.e. $\sim 4$ Gyrs) or old GCs with blue horizontal branches (hereafter young/BHB GCs). Figure \ref{fig:C07data} shows that the B00 relationship between color and metallicity (derived for Galactic GCs) is also consistent with the NGC 1407 GC data. Noticeable in Figure \ref{fig:C07data} is the position of the 3 young/BHB GCs, which agrees with the 5 Gyrs V03 model line. However, as mentioned in C07, their position could also be explained by the presence of a blue horizontal branch (BHB) in these clusters.

\begin{figure}
\epsscale{1.19}
\plotone{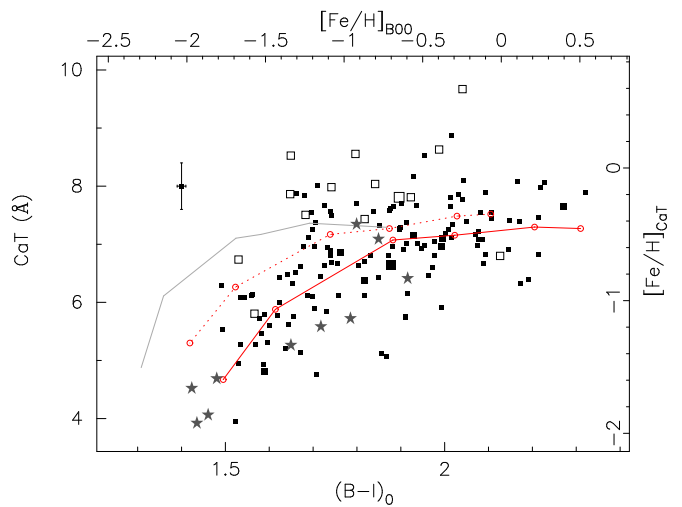}
\caption{The distribution of NGC 1407 GCs in the $(B-I)_{0}$-CaT plane. Solid and dotted red lines are the V03 models for 13 and 5 Gyr, respectively. Young models of age 1.6 Gyrs are also plotted as a solid grey line. Grey stars are Galactic GCs ($E(B-V) \le 0.3$) with colors taken from H96 and CaT values from AZ88 (after applying the inverse of Equation \ref{eq:convCaT}). Squares show our data with relative sizes proportional to the the signal-to-noise in the raw spectrum. Hollow symbols are discussed in Appendix \ref{sec:SmallLines}. The top x-axis shows the B00 color-metallicity relationship based on Galactic GCs and the right y-axis shows the metallicity derived from CaT using Equation \ref{eq:convtoFe}. Typical error bars shown.}\label{fig:BICaT}
\end{figure}

The results presented in Figure \ref{fig:AZ88}, \ref{fig:compare} and \ref{fig:C07data} suggest that, under the assumption that the GCs in NGC 1407 are not significantly different from those of the Milky Way (i.e. similar stellar content), the CaT index values should scale linearly with $(B-I)_{0}$ colors (at least for metallicities below [Fe/H]$\lesssim-0.5$ according to the V03 models). Figure \ref{fig:BICaT} shows the relationship between the $(B-I)_{0}$ colors and CaT for our sample of GCs. The CaT values for the Milky Way GCs were converted from the CaT$_\mathrm{AZ88}$ quoted in AZ88. The agreement between the NGC 1407 dataset and the V03 models is good with the data scattering about the 13 Gyr track. While a general trend between color and CaT is observed, there are several interesting features present. Indeed, it appears as though the CaT index flattens out or `saturates' for metallicities higher than about [Fe/H]$\approx-0.5$ (on the right y-axis) as predicted by V03. This behavior is in contradiction with the prediction of increasing metallicity sensitivity made by BC03 and with an extrapolation of the AZ88 relation to higher metallicities. We therefore decided to consider the V03 models more closely than the BC03 models as they yield better agreement.

The bulk of the GCs around NGC 1407 (and the V03 models) apparently have either higher CaT index values, or bluer colors, than the Galactic GCs as shown in Figure \ref{fig:BICaT}. This apparent offset between the Galactic GCs and the NGC 1407 GCs could be related to the lack of flux calibration in both this work and AZ88, which makes the two studies difficult to compare directly. Moreover, the small number and lack of redder Galactic GCs complicates this comparison. Also, the Galactic GCs are plagued by large photometric uncertainties due to Galactic extinction. Indeed, Figure 14 of V03 has already shown that the behavior of the CaT with metallicity in Galactic GCs is in reasonable agreement with their models. Because of the uncertainties involved, the relative position of the Galactic GCs along both axes of Figure \ref{fig:BICaT} is uncertain and we refrain from drawing strong conclusions from it.

\begin{figure}
\epsscale{1.19}
\plotone{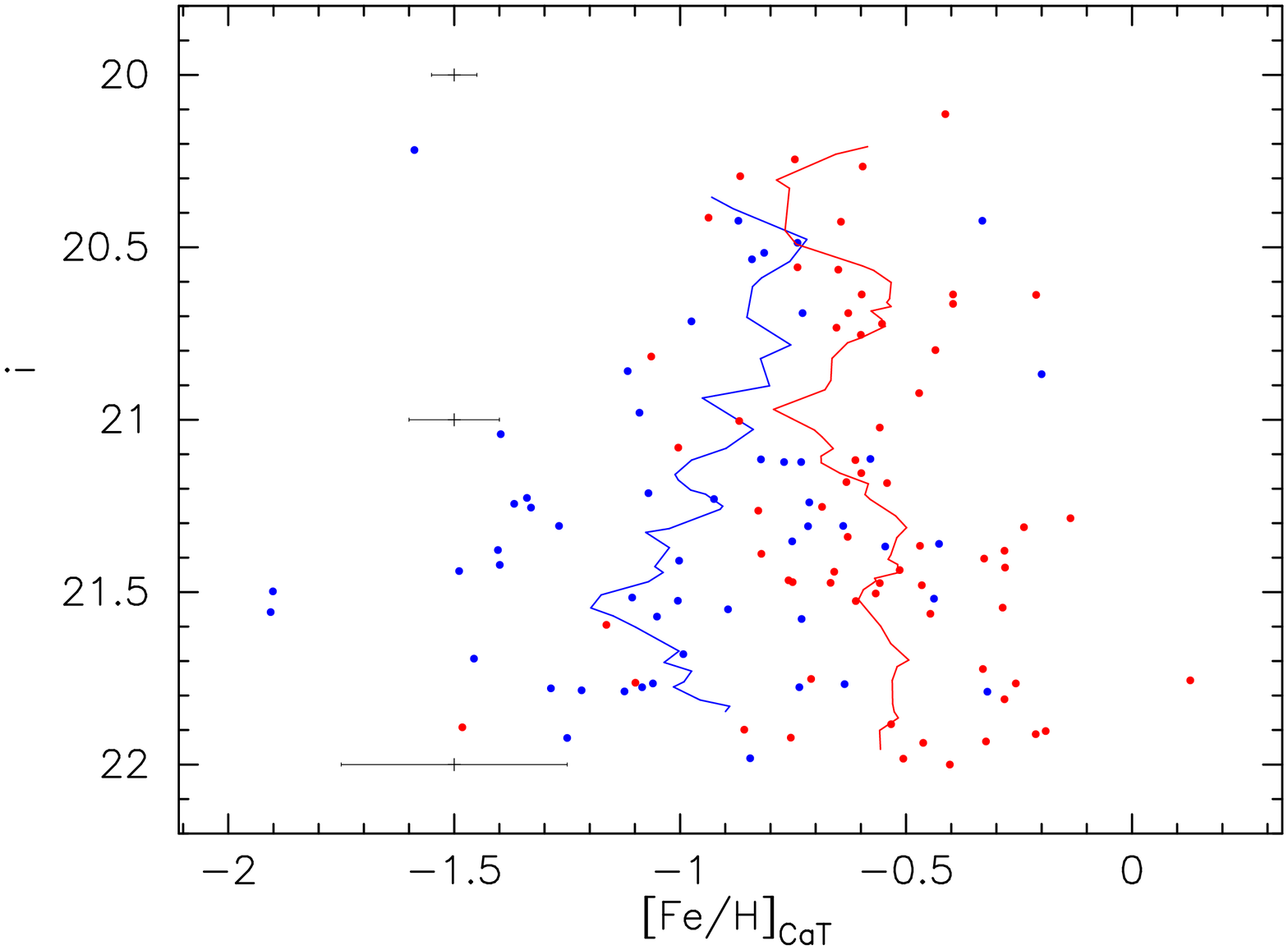}
\caption{The inferred metallicity distribution for our sample of NGC 1407 GCs. A running average (solid blue and red lines) have been overplotted for the blue and red subpopulations (colors are as inferred from the photometry). Typical error bars are shown (see Appendix \ref{Appendix:errors}).}\label{fig:tilts}
\end{figure}

\begin{figure}
\epsscale{1.19}
\plotone{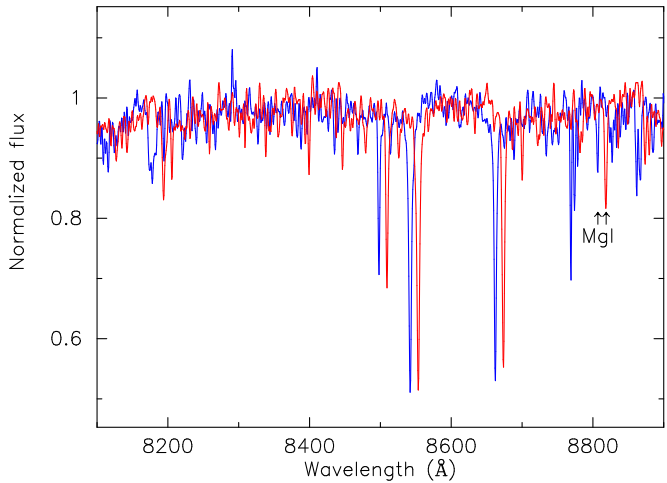}
\caption{Averaged and continuum normalized raw spectra of the brightest ($i\le20.5$) blue and red GCs in NGC 1407. The red spectrum has been shifted by 200 km s$^{-1}$ from a zero redshift for clarity. Even though their average $(g-i)_{0}$ colors are separated by over 0.2 mag, the CaT line strengths of the two mean raw spectra are nearly identical suggesting similar metallicities. This is in contrast with the MgI feature that is deeper in the red GCs as expected for higher metallicity.}\label{fig:brightGC}
\end{figure}

The GC metallicity distribution obtained by applying our CaT to [Fe/H]$_{\mathrm{CaT}}$ transformation (Equation \ref{eq:convtoFe}) is shown in Figure \ref{fig:tilts}. There are several unexpected features present in the inferred metallicity distribution. First, the blue tilt observed in the CMD (Figure \ref{fig:cmd}) appears to still be present, however there is also a hint of an inverse red tilt. The spread in metallicity of the blue GCs seems to be wider than that of the red GC subpopulation in contrast to their spreads in color that are similar for both subpopulations. Moreover, using the CaT as a metallicity indicator seems to have merged the red and blue bright peaks even though they are separated in color by more than $\Delta (g-i)_{0} \approx 0.2$ or $\Delta (B-I)_{0} \approx 0.25$ mag. This color separation corresponds to an expected metallicity difference of over 0.8 dex for old ages (V03). Figure \ref{fig:brightGC} shows the mean raw spectra of the brightest ($i<20.5$) red and blue GCs. They have similar CaT line strengths and hence inferred metallicity. The same exercise was repeated using the median value of the brightest spectra with, and without fitting, to ensure that the average was not biased by one outlier spectrum, skylines or our template fitting procedure. Each time there was no obvious difference between the strength of the CaT features in the bright red and blue GC spectra.

On the other hand, the averaging of the brightest red GCs has increased the signal-to-noise ratio sufficiently to detect the Mg\thinspace{\sc{i}} line at 8807\AA\space while this feature is barely present in the mean spectrum of the brightest blue GCs. We measure the generic Mg\thinspace{\sc{i}} index defined in Cenarro et al. (2009) for both spectra and find that the brightest red GCs have Mg\thinspace{\sc{i}}$=0.77\pm$0.13\AA\space while the brightest blue GCs have Mg\thinspace{\sc{i}}$=0.15\pm$0.15\AA.
The Mg\thinspace{\sc{i}} line is mostly sensitive to both temperature and [Mg/H] which correlates with [Fe/H] (see Cenarro et al. 2009). This suggests that the bright blue GCs are either more typical of a hotter (earlier-type) or a less metal enriched stellar population than the red ones. The latter is more likely since hot stars contribute very little at near-infrared wavelengths.

\begin{figure}
\epsscale{1.19}
\plotone{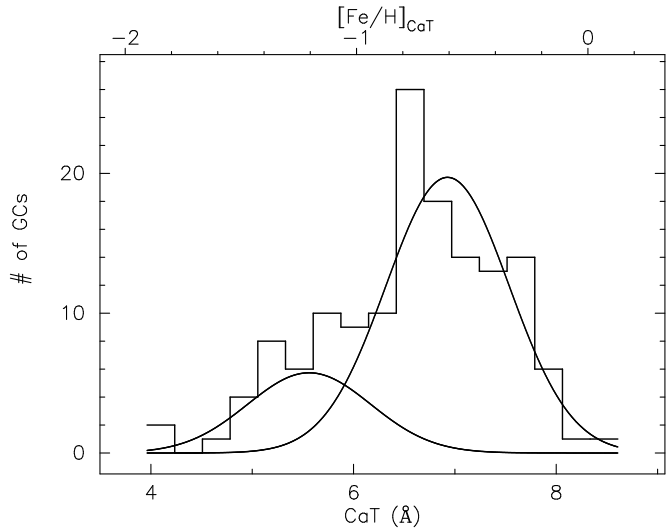}
\caption{A histogram showing the distribution of the CaT (lower axis) and [Fe/H]$_{\mathrm{CaT}}$ (upper axis) in our sample. The KMM test shows that the distribution of our data is best fitted with two Gaussians (overlaid) at the 96 \% confidence level. The inferred mean metallicities are [Fe/H]$_{\mathrm{CaT}}=-1.20$ and [Fe/H]$_{\mathrm{CaT}}=-0.61$ for the blue and red subpopulations, respectively.}\label{fig:histogram}
\end{figure}

At fainter magnitudes a difference in CaT is seen. The KMM test (Ashman, Bird \& Zepf 1994) performed on the CaT inferred metallicities for the \emph{whole sample} shows that a bimodal distribution is preferred over a unimodal one for this dataset at the 96 \% confidence level. The output mean metallicities are [Fe/H]$_{\mathrm{CaT}}=-1.20$ and [Fe/H]$_{\mathrm{CaT}}=-0.61$ for the blue and red subpopulations, respectively. This can arguably be seen in the non-symmetric shape of the [Fe/H] or CaT values histogram (see Figure \ref{fig:histogram}). We are however careful drawing any strong conclusions from these results as: 1) the shape of the distribution seen in Figure \ref{fig:histogram} is considerably different from that seen in the color histogram (Figure \ref{fig:BIhist}), and 2) the numbers are low. Moreover, the confidence level drops below significance (87 \% only) when the KMM test is performed on the CaT$_{\rm AZ88}$ index measured on the raw data possibly due to the larger measurement errors.  Nevertheless, with these caveats in mind, we compare with Forbes et al. (2006) who found an average of [Fe/H]$ = -1.45$ and [Fe/H]$ = -0.19$ for the metal-poor and -rich GC subpopulations based on $(B-I)_{0}$ photometry for NGC 1407 GCs. The inconsistently lower mean metallicity for the red subpopulation than that found in the analysis by Forbes et al. (2006) could be explained by: 1) possible radial gradients (see Section \ref{sec:data}), and/or 2) the prediction of V03 of the saturation of the CaT feature around [Fe/H]$\sim-0.5$. However, it appears that the blue GC subpopulation has systematically higher measured CaT than expected. This cannot be attributed to either radial trends or saturation effects. It is puzzling that the absolute position of the blue GCs is shifted towards higher metallicities causing the two subpopulations' metallicity distributions to be similar for the most luminous GCs.

\section{Discussion}\label{sec:discussion}

Based on the results presented in Figures \ref{fig:AZ88}, \ref{fig:compare}, and \ref{fig:C07data}, the CaT index should scale linearly with metallicity and $(B-I)_{0}$ color at least for [Fe/H]$\le-0.4$. On the other hand, single stellar population models disagree about the sensitivity and behavior of the CaT features with respect to metallicity. The empirical calibration of AZ88 was adopted throughout this work to derive metallicities from the CaT index values under the assumption that the GCs around NGC 1407 are intrinsically similar to those around the Galaxy.

As demonstrated in Figure \ref{fig:brightGC}, the bright blue and red GCs in NGC 1407 have the same CaT index values. This would suggest that they have the same average metallicity even though they differ in mean $(g-i)_0$ color by over 0.2 mag. The metallicity inferred from the AZ88 relationship for both the brightest blue and red GCs is [Fe/H]$_{\mathrm{CaT}}\sim-0.8$ dex.

Moreover, even though the GC system is clearly bimodal in color and in CaT, which is measured on the fitted spectra, it is not bimodal in the CaT$_{\rm AZ88}$ index distribution measured on the raw spectra. This is possibly due to increased measurement errors of the indices measured on the raw spectra. For this reason, we do not consider that our CaT data are sufficient to confidently confirm (i.e. with more than 95 \% confidence) that the GC system around NGC 1407 is bimodal in CaT inferred metallicity contrary to what is expected from the colors. However, in Appendix \ref{sec:SmallLines} we show how the distribution of the sum of the equivalent width of three weak features present in the \emph{fitted spectra only} exhibit clear bimodality.

The disagreement between the metallicities inferred from the colors and those inferred from the CaT index values points to a fundamental difference between the GCs around NGC 1407 and the Milky Way either with respect to the behavior of their colors or the CaT index with metallicity.

However, it is possible that the data have some systematic biases that could cause this behavior. Such a systematic effect would need to affect the blue GC subpopulation much more than the red one. One possible source of systematics is the sky subtraction. If the sky was oversubtracted for the blue GCs only, this would yield higher CaT measurements. This was investigated and there is no obvious offset between the continuum levels of the blue and red subpopulations. The behavior of the CaT with decreasing signal-to-noise was tested by adding Poisson noise to V03 SSP models SEDs of various metallicities. While lower signal-to-noise spectra inevitably yield to larger errors on the measured CaT index values, no systematic scattering to higher or lower values, or with metallicity or color was found (see Appendix \ref{Appendix:errors}).

Assuming the data are reliable we now explore several possible explanations. One possible explanation to the similar metallicities inferred for the bright blue and red GCs is that the CaT features saturate at lower metallicities than predicted by the SSP models (i.e. at [Fe/H]$_{\mathrm{CaT}}\sim-0.8$ instead of [Fe/H]$_{\mathrm{CaT}}\sim-0.5$). If we make the reasonable assumption that our composite near-infrared spectra are not significantly influenced by hot early-type stars, the presence of a stronger Mg\thinspace{\sc{i}} line in the mean spectrum of the bright red GCs than in that of the bright blue GCs does indeed suggest that the CaT could be saturated while the Mg\thinspace{\sc{i}} is still sensitive to metallicity. However, the data presented in Figure \ref{fig:BICaT} suggests a saturation metallicity that is in agreement with the V03 models. Nevertheless, early saturation of the CaT could happen if the GCs around NGC 1407 were Calcium enhanced with respect to their Galactic counterparts (i.e. [Ca/Fe] is larger in NGC 1407 GCs). Indeed, the data presented in Battaglia et al. (2008) for the CaT in individual red giant branch stars suggests that a different [Ca/Fe] ratio than is present in the calibration GCs may influence the measured CaT values and thus the inferred metallicities by up to about 0.2 dex error for [Fe/H].

A saturation of the CaT features at lower metallicities could also be the result of a more bottom-heavy IMF for the GCs around NGC 1407 with respect to those around the Galaxy. Because the CaT index decreases with increasing surface gravity (DTT89; Cenarro et al. 2002), dwarf stars have lower CaT values. A stellar population with a bottom-heavy IMF would contain a larger proportion of dwarf stars at old ages and thus should saturate at lower CaT index compared to a more top-heavy IMF. For example, models with a Salpeter (1955) IMF saturate at higher CaT values than models with a Kroupa (2001) IMF in the V03 SSP models. The Mg\thinspace{\sc{i}} index is only minimally influenced by gravity (Cenarro et al. 2009), so it is unlikely to be significantly influenced by a different IMF and should still be a good tracer of metallicity.

The relative spread in CaT inferred metallicity of the blue and red subpopulations are at odds with what is expected from their respective spread in color. If proven correct, this could be due to the non-linearity of the conversion between colors and metallicity (e.g.,Yoon et al. 2006; Peng et al. 2006). In some galaxies, a non-linear conversion between color and metallicity could cause a unimodal metallicity distribution to appear bimodal in color (Cantiello \& Blakeslee 2007; Blakeslee et al. 2010). Therefore, if we take our inferred metallicities at face value, the color and metallicity distributions could be reconciled by invoking this effect.

Alternatively, the difference in the relative spreads of the respective subpopulations between the color and CaT distributions could be a result of systematic biases. Indeed, the higher sensitivity of the CaT to metallicity for metal-poor (blue) GCs, which have metallicities lower than the saturation limit predicted by V03, allows for a wider range of inferred metallicities. However, the saturation limit is reached at metallicities corresponding to the metal-rich (red) GCs. This in turn reduces the range of allowed CaT values and thus the inferred range in metallicities for the red GC subpopulation.

It is possible, although an extreme case, that the majority of our blue spectra are contaminated by Paschen line absorption. If this were true it could play a definite role in explaining our inferred metallicity distribution. There are several spectral features of the Paschen series of hydrogen in the spectral region of the CaT. Their individual depths vary slightly with the deepest lines at redder wavelengths. Because three of the features of the Paschen series of Hydrogen overlap with the three CaT features, even a small level of contamination or order $\approx 0.3$\AA\space per overlapping Paschen line would be sufficient to significantly alter the measured CaT index by $\approx0.9$\AA. Paschen lines are predominant in stellar spectra of B, A, and F spectral types. These usually massive hot stars are short lived and thus old stellar populations such as GCs are not expected to show significant Paschen absorption features from such stars. Indeed, Paschen lines only become significant in the V03 SSP models for ages $\lesssim2.0$ Gyrs and metallicities [Fe/H]$\lesssim-0.68$. However, keeping in mind that models are uncertain at young ages (V03), we cannot rule out the possibility these GCs could be young. Non-overlapping Paschen lines around the CaT (i.e. Pa12, Pa14 and Pa17) with equivalent widths of order $\approx 0.3$\AA\space would not be detectable in our modest signal-to-noise spectra. However, in Appendix \ref{sec:SmallLines} we discuss how Paschen lines are sometimes visible in the \emph{fitted spectra only}.

Another alternative is a population of GCs with hot blue stars such as BHB or blue stragglers stars (Cenarro et al. 2008) at intermediate metallicities. This is consistent with the favored conclusion of C07 regarding the young/BHB GCs in NGC 1407 based on the diagnostic of Schiavon et al. (2004). Moreover, there is some evidence from UV studies that the GC systems of massive elliptical galaxies such as NGC 1407 could harbour a significant population of GCs with extreme hot horizontal branches (see Sohn et al. 2006; Mieske et al. 2008). We examined the archival \emph{Galaxy Evolution Explorer} (GALEX) UV images of NGC 1407 and found that they were not deep enough for its GC system to be detected. If an additional contribution from Paschen line to the CaT index caused by hot blue stars is indeed present, then NGC 1407 could harbour a population of \emph{intermediate metallicity} GCs with BHBs or extreme HBs as speculated from UV studies of other galaxies or a population of GCs with a large proportion of blue straggler stars.

With this in mind, it is worth reconsidering Figure \ref{fig:C07data} and the results of C07. The $(r-i)_{0}$ colors should be less affected by, although not immune to, the presence of hot blue stars. This is because hot blue stars contribute less of the integrated light at redder wavelengths (e.g.,Smith \& Strader 2007; Spitler, Forbes \& Beasley 2008; Cantiello \& Blakeslee 2007). We therefore compare the position of the 3 young/BHB GCs identified by C07 in both color spaces. The position of the 3 young/BHB GCs are consistent with the bulk of the other GCs with measured old ages using $(r-i)_0$ colors (right hand panel of Figure \ref{fig:C07data}).

In summary, the unexpected distribution of CaT values and inferred metallicities for NGC 1407 GCs could be explained by 1) early saturation of the CaT features in NGC 1407's GCs, 2) a population of GCs with hot blue stars in NGC 1407, and/or 3) the non-linear conversion between color and metallicity. With the current dataset and based on the current generation of SSP models at near-infrared wavelengths, we cannot positively determine which (combination) of these possible explanations is the correct one. Unfortunately, this casts serious doubts on the CaT inferred metallicity distribution shown in Figure \ref{fig:tilts}, the presence of a spectroscopic blue tilt, and the potential of the NIR CaT feature as a metallicity indicator in the integrated light spectra of extragalactic GCs. Until these issues are understood, metallicities inferred from the CaT for integrated light spectroscopy of extragalactic GCs will remain uncertain.

\section{Summary and Future work}\label{sec:conclusions}

The empirical relationship found by Armandroff \& Zinn (1988) between the CaT$_\mathrm{AZ88}$ index values and metallicities in Galactic GCs was revisited in the light of more recent metallicity measurements. No noticeable difference was found and based on the literature we conclude that the CaT can in principle be used to determine metallicities in integrated spectroscopy of extragalactic GCs. This assumes that the Galactic GCs are not intrinsically different from extragalactic GCs.

We also compare the predictions for the behavior of the CaT with metallicity for the V03 and BC03 SSP models. We find that different SSP model sets assign widely different absolute metallicities for a given CaT value.

A sample of 144 GCs in NGC 1407 suitable for stellar population analysis were obtained in the spectral region near the CaT using DEIMOS on Keck. The metallicity distribution for this sample was obtained based on the empirically determined conversion of AZ88 from CaT$_\mathrm{AZ88}$ index values. Several unexpected results were obtained, the most notable of which is the identical CaT index values for the brightest blue and red GCs. Even though the bright red and blue GCs are well separated in color space, the CaT measurements suggest that they have a similar metallicity. We show that this result is independent of the index measurement method used since the average raw spectra for the brightest blue and red GCs themselves are nearly identical.

Integrated light spectra of nearby (Local Group) resolved GCs with independently measured ages, metallicities and HB morphologies determined via color-magnitude diagrams are essential to determine what effects metallicity, age, HB stars or blue straggler stars have on the integrated light spectra of GCs in the near-infrared. Such a dataset would also help our understanding of these phenomena and may allow for the calibration of the CaT as a metallicity indicator for extragalactic GCs.

\section*{Acknowledgements}
We thank the referee for his/her constructive comments. We would like to thank Javier Cenarro, Alan Brito, George Hau and Aaron Romanowsky for helpful insight and discussions. We are grateful to J. Trevor Mendel for help with the template fitting procedure. CF thanks the Anglo-Australian Observatory for financial support in the form of a graduate top-up scholarship. DAF and RNP thank the ARC for financial support. This material is based upon work supported by the National Science Foundation under Grant AST-0507729. Support for this work was provided by NASA through Hubble Fellowship grant \#HF-01214.01 awarded by Space Telescope Science Institute, which is operated by the Association of Universities for Research in Astronomy, Inc., for NASA, under contract NAS 5-26555. The data presented herein were obtained at the W.M. Keck Observatory, which is operated as a scientific partnership among the California Institute of Technology, the University of California and the National Aeronautics and Space Administration. The Observatory was made possible by the generous financial support of the W.M. Keck Foundation. The analysis pipeline used to reduce the DEIMOS data was developed at UC Berkeley with support from NSF grant AST-0071048. Based in part on data collected at Subaru Telescope, which is operated by the National Astronomical Observatory of Japan.

\begin{appendix}

\section{Error determination and sample selection}\label{Appendix:errors}

We use the method described in Cardiel et al. (1998) and in appendix A2 of C01 to compute the errors on the raw indices (CaT$_{\rm AZ88}$). For the CaT index measured on the template fitted spectra it is difficult to assess the errors in our measurements and hence on the inferred metallicity exactly. This is because the template fitting procedure has modified the variance on each pixel in a nontrivial manner. While the fitting has essentially extrapolated across skylines, thereby minimizing their effect on the final index measurements, it is unclear how the fits themselves may be affected by skyline residuals and noise. Adding to this is the possible introduction of errors arising from the continuum fitting step.

\begin{figure}
\epsscale{0.7}
\plotone{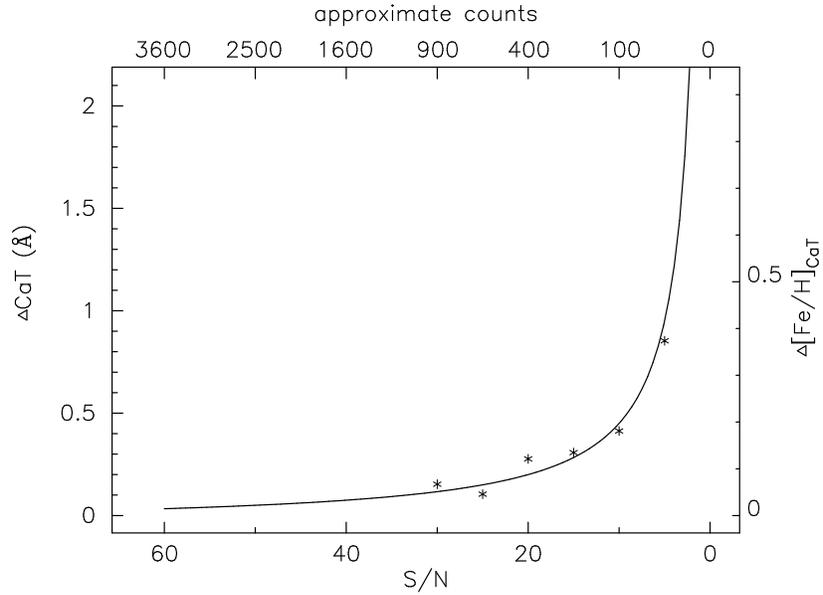}
\caption{Errors on the CaT and [Fe/H] as a function of signal-to-noise per \AA\space (S/N). The stars show the average difference between the CaT measured directly on the SSP model spectrum and that measured on the same spectrum but with added Poisson noise. The solid line is a fit to these data points. The top abcissa shows the number counts expected if the spectra contain Poisson noise only. Our observations range from $9\le$S/N$\lesssim60$}\label{fig:SN_deltaCaT}
\end{figure}

For this reason, we model the error on our CaT measurements using Monte Carlo methods similar to that used by Emsellem et al. (2004) and Weijmans et al. (2009). We first measure the `exact' CaT index value using our index measurement method (template and continuum fitting applied) on V03 SSP model spectra. These are then compared to CaT values measured on the same fitted spectra with added Poisson noise. The errors as a function of signal-to-noise shown in Figure \ref{fig:SN_deltaCaT} are the average difference between the two measurements. We find no trend between the errors and metallicity.

The method detailed above does not describe our CaT errors perfectly. Nevertheless, it follows our index computation scheme closely and should be a reasonable error estimate. This is supported by the fact that the scatter in the CaT and [Fe/H] distributions is similar to the size of the error bars. There is one caveat with these error estimates: the effect of skyline residuals and other non-Poisson noise sources are not included. We suspect the presence of skyline residuals have a small effect on the errors since skyline contaminated regions are excluded by the \textsc{pPXF} routine.

\begin{figure}
\epsscale{0.7}
\plotone{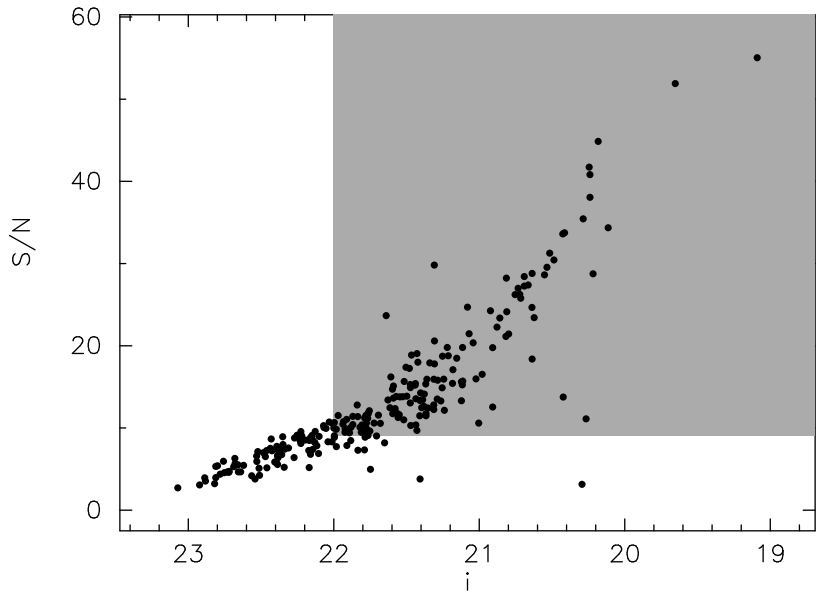}
\caption{Variation of the average signal-to-noise per \AA\space (S/N) as a function of $i$-band magnitude for the spectroscopically confirmed GCs. The shaded grey area shows the sample of 144 GCs selected for the CaT analysis.}\label{fig:i_SN}
\end{figure}

Based on the size of the errors we removed spectra with raw counts less than 80 (S/N$\sim9$ \AA$^{-1}$) from our sample, which corresponds to roughly an $i$-band magnitude of 22.0 (see Figure \ref{fig:i_SN}). This yields a maximum CaT error of 0.5 \AA \space and a maximum [Fe/H] error of roughly 0.25 dex. This left a sample of 144 spectra, sufficient for a statistical analysis.

\section{Other weak features}\label{sec:SmallLines}

In this section, we explore other weak spectral features that are not clearly visible in the raw spectra but can be measured in the fitted spectra only. While the relative strength of these weak lines may be an artifact of the fitting technique, we find that there are some interesting characteristics that may help interpret our measured CaT distribution.

The first interesting feature is the presence of the Pa12 line in a non-negligible number of our fitted spectra. An example of this is seen in Figure \ref{fig:indices}. It is the broad feature centered at 8751 \AA\space and it is overlaid by a multitude of smaller features. The Pa12 line is part of the Paschen series of which 3 features overlap with the CaT features, namely the Pa13, Pa15, and Pa16 lines.

SSP model spectra do not show such Paschen lines with corresponding intermediate colors for ages $\gtrsim2.0$ Gyrs although one must keep in mind the increased uncertainty of the models at young ages. The hollow symbols in Figure \ref{fig:BICaT} show the GCs with measured Pa12 $\ge1.65$ \AA. Most of the data that are scattered to higher CaT index values are those with possibly large Pa12 absorption. Indeed, roughly 20 \% of the fitted spectra at intermediate colors/metallicity have Pa12 $\ge1.65$. The inferred average Pa12 index value for the bright blue GCs is 0.3 \AA\space greater than that of the bright red GCs indicating that the blue GCs may contain a higher degree of Paschen contamination. Unfortunately, it is not possible with the current data to positively confirm the presence of Paschen line contamination.

We find that our fitted spectra also contain tentative evidence for the presence of several other `weak lines' (i.e. Ti\thinspace\textsc{i} $\lambda$8436\AA, Fe\thinspace\textsc{i} $\lambda$8514\AA\space and Fe\thinspace\textsc{i} $\lambda$8689\AA, see Figure \ref{fig:indices}). As with the Pa12 feature, these are not seen in our raw spectra but they appear in the fitted spectra. These lines are metallicity sensitive lines and could thus serve as potential metallicity indicators provided an appropriate calibration and signal-to-noise. As a matter of fact, other groups have made use of various spectral features in the region of the CaT to measure stellar metallicities using high signal-to-noise DEIMOS spectra (e.g.,Kirby et al. 2008; Shetrone et al. 2009).

\begin{figure}
\epsscale{0.7}
\plotone{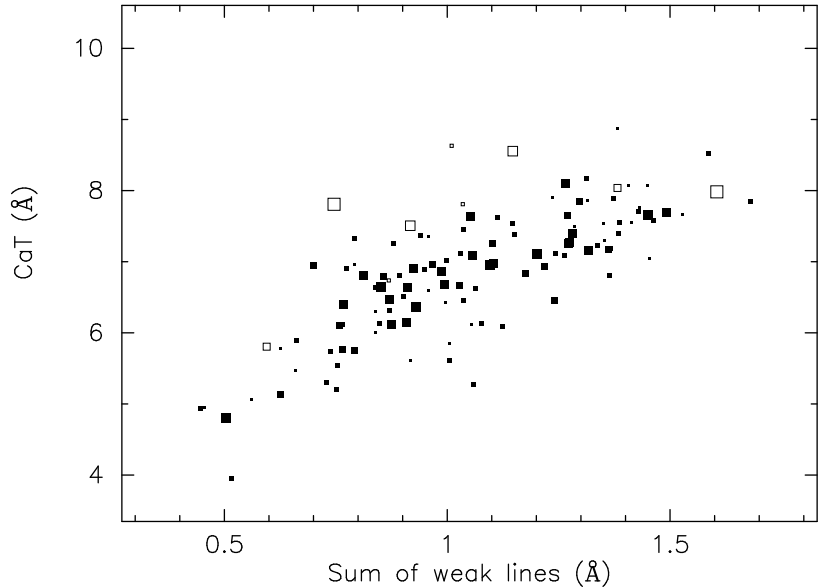}
\caption{Correlation between the CaT and the sum of the equivalent widths of the weak lines (Ti\thinspace\textsc{i} $\lambda$8436\AA, Fe\thinspace\textsc{i} $\lambda$8514\AA\space and Fe\thinspace\textsc{i} $\lambda$8689\AA), which are likely sensitive to metallicity. Filled and hollow squares correspond to Pa12 $< 1.65$ \AA\space and Pa12 $\ge 1.65$ \AA, respectively, with relative sizes proportional to the the signal-to-noise in the raw spectrum. The overall trend suggests that both CaT and the weak lines are tracing metallicity.}\label{fig:small_CaT}
\end{figure}

\begin{figure}
\epsscale{0.7}
\plotone{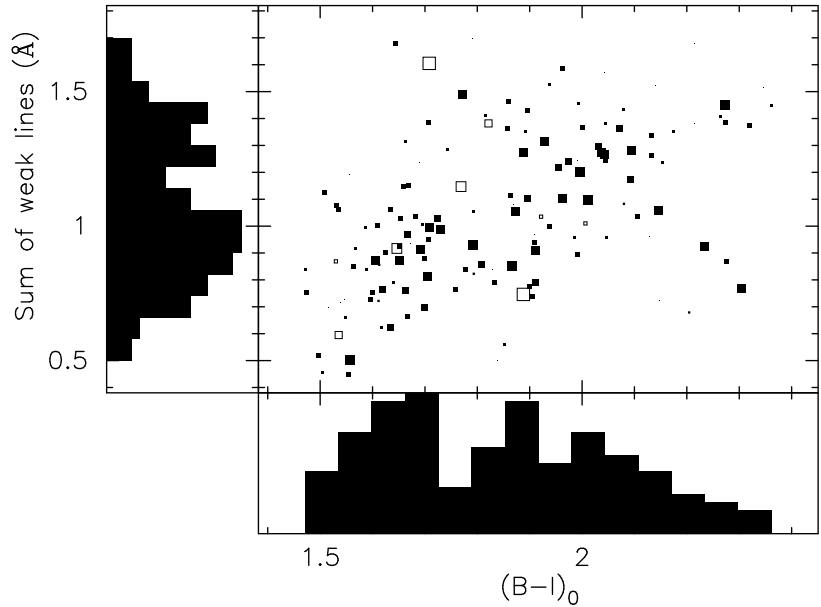}
\caption{Correlation between $(B-I)_0$ color and the sum of the equivalent widths of the three weak lines. Histograms of each variable show that both color and the equivalent widths of the weak lines are bimodal.}\label{fig:BI_small}
\end{figure}

Figures \ref{fig:small_CaT} and \ref{fig:BI_small} show that the sum of the equivalent widths of the weak lines correlate with both CaT and color albeit with large scatter. The tighter correlation between the weak lines and CaT (Figure \ref{fig:small_CaT}) suggests that both trace metallicity similarly. On the other hand, while bimodality is not readily visible in the distribution of CaT, it is clear what the distribution of the sum of the weak lines is bimodal. Indeed, the KMM test concludes that a bimodal distribution is preferred over a unimodal at the 97 \% confidence with 65 and 35 \% of the GCs being blue and red, respectively. This is in contrast to their color proportion that are 40 and 60 \% for the blue and red GCs, respectively.

When fitting our spectra we do not directly fit the small lines, which are not seen in the raw spectra. The templates themselves contain information about the relative ratios of these lines to CaT for a given metallicity such that the distribution for the sum of the weak lines displays the expected bimodality. We speculate that while the CaT features may be saturated at high metallicity, the weak lines could still be on the growth curve and their sum a better proxy for metallicity.

We are very cautious however as these features are solely measured as a result of the template fitting method and thus the inferred bimodality could be an artifact. On the other hand, if it were an artifact it is not clear how the fitting method could conspire to create a clearer (i.e. with higher confidence) bimodality than seen in the CaT distribution.

\end{appendix}

\end{document}